\documentclass[11pt,a4paper]{article} \usepackage{jheppub}

\title{On the next-to-leading order QCD 
${\cal K}$-factor for  $t\bar{t}b\bar{b}$ 
production at the TeVatron}

\author[a]{Ma\l{}gorzata Worek} 
\affiliation[a]{Fachbereich C, Bergische Universit\"at Wuppertal,  
D-42097 Wuppertal,    Germany} 
\emailAdd{worek@physik.uni-wuppertal.de} 

\abstract{ Motivated by ongoing experimental analyses, we report on
  the calculation of next-to-leading order
  QCD corrections to  the production of $t\bar{t}$ pairs in
  association with two hard b-jets at the Fermilab  TeVatron.  Besides
  the total cross section and its scale dependence, a few
  differential distributions applicable for Higgs boson searches at
  the TeVaron are given. The QCD corrections with respect to leading
  order are negative and small. For our main setup they amount to
  $2\%$, and remain reasonably stable against changes of cuts. This
  proves that an 
  integrated  next-to-leading order ${\cal K}$-factor does not necessarily
  need to be applied in the background estimation for the
  $t\bar{t}H\to t\bar{t}b\bar{b}$ signal process. The distributions
  show similarly small corrections.  The shape of kinematic
  distributions is distorted at most by about $20\%$ in some
  corners of the phase space.   Even though it is not the main purpose
  of this paper, we also evaluated the forward-backward asymmetry of the
  top quark at  next-to-leading order. } 

\dedicated{\rm WUB/11-25}

\begin{document} 
\maketitle
\flushbottom


\section{Introduction}

%
The production of the Standard Model Higgs boson in association with a 
top-anti-top pair allows for a direct study of the top  Yukawa
coupling, which is  an important  step in understanding the nature of
the electroweak symmetry breaking mechanism. The 
$t\bar{t}H$ production  channel is especially important in the range 
of Higgs boson masses
$m_H \le 140 ~{\rm GeV}$, where the Higgs boson decays
predominantly into $b\bar{b}$ pairs. While the LHC is making ground
breaking progress towards discovering the Higgs boson, and certainly
tightens exclusion bounds well past those obtained at the TeVatron,
the analysis of the data from the latter is still being completed. It
is a fact that the predicted cross section for
Higgs boson radiation off top quarks at the TeVatron is rather
low \cite{hep-ph/0107081,hep-ph/0107101,hep-ph/0109066,hep-ph/0211352}. 
Therefore,  a discovery of the Higgs boson in this channel alone 
would not have been possible. However, it contributes to
the combination of the Standard Model Higgs 
boson searches \cite{arXiv:1107.5518}. 
The main background process consists of direct production of the final 
state without resonances, i.e. of the QCD generated process
$t\bar{t}b\bar{b}$. The procedure used by the experimental
collaborations for the background estimate, 
is either direct use of leading order (LO) Monte Carlo
simulations \cite{D0}, or additional reweigthing by a K-factor
\cite{CDF}. Unfortunately, a next-to-leading order (NLO)  K-factor for this
process under TeVatron conditions has never been published. According
to \cite{CDF}, the analysis employs the published value
\cite{arXiv:0907.4723} for the LHC. Since the production mechanisms
are very different  in
both cases, it is questionable that such a procedure
leads to reliable estimates. In this paper, we address this issue by
providing an NLO QCD prediction to
$t\bar{t}b\bar{b}$, the  irreducible QCD background to the
$t\bar{t}H\to t\bar{t}b\bar{b}$  process at the TeVatron.
%

\section{Theoretical Framework}

\begin{figure}
\begin{center}
\includegraphics[width=0.95\textwidth]{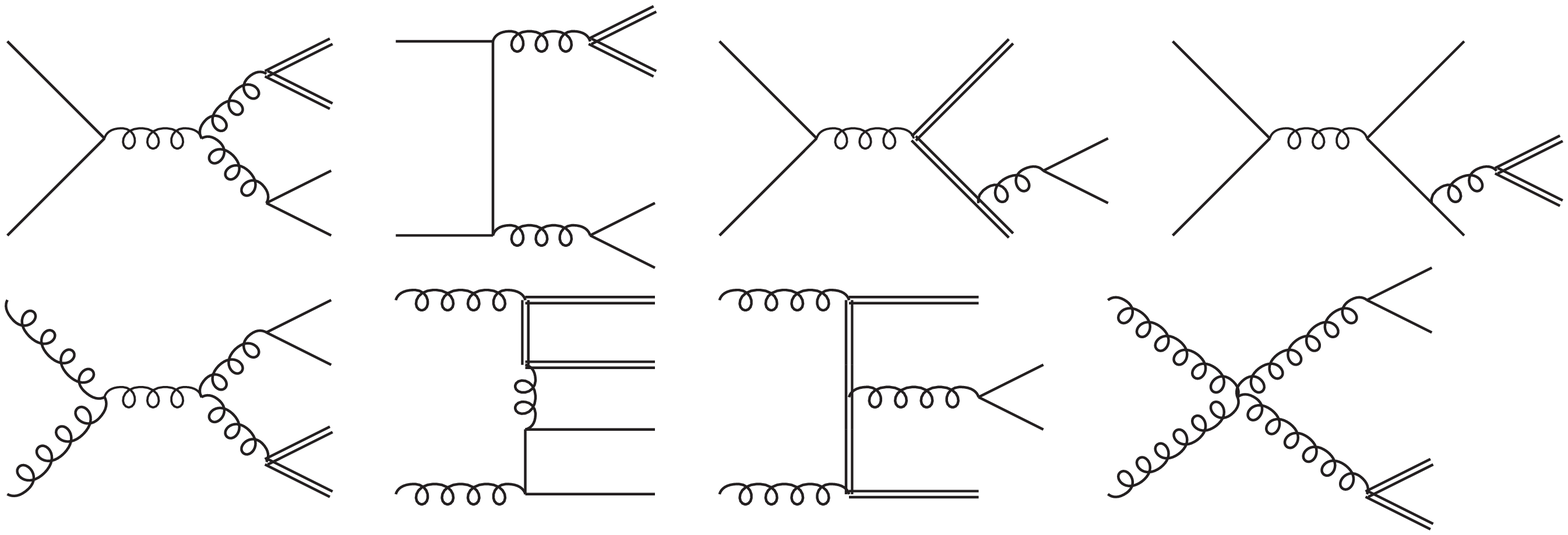}
\end{center}
\vspace{-0.2cm}
\caption{\it \label{fig:LO}  A representative set of Feynman diagrams
  contributing to the leading order  $p\bar{p} \to t\bar{t}b\bar{b}$ 
  process. Double lines correspond to top quarks,
  single lines to light quarks and wiggly ones to gluons.}
\end{figure}

\begin{figure}
\begin{center}
\includegraphics[width=0.95\textwidth]{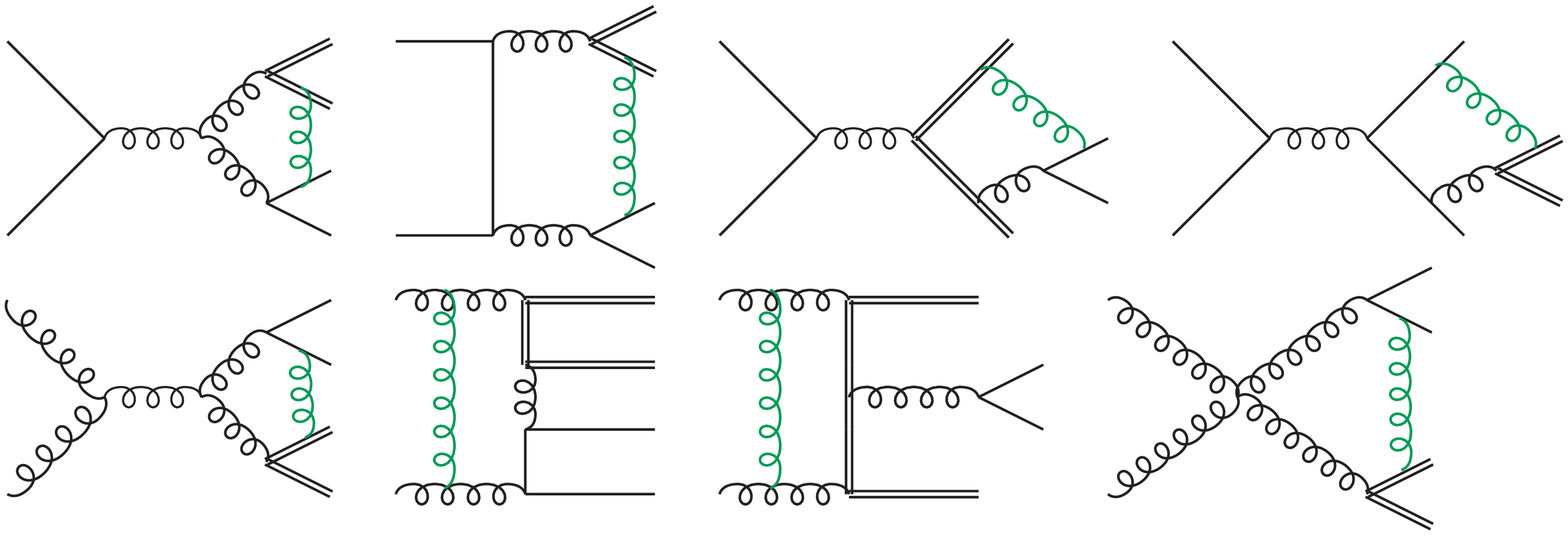}
\end{center}
\vspace{-0.2cm}
\caption{\it \label{fig:V}  A representative set of pentagon and hexagon
  diagrams contributing to the virtual corrections to the $p\bar{p} \to
  t\bar{t}b\bar{b}$  process. Double lines
  correspond to top quarks, single lines to light quarks and wiggly ones to
  gluons.}
\end{figure}

\begin{figure}
\begin{center}
\includegraphics[width=0.95\textwidth]{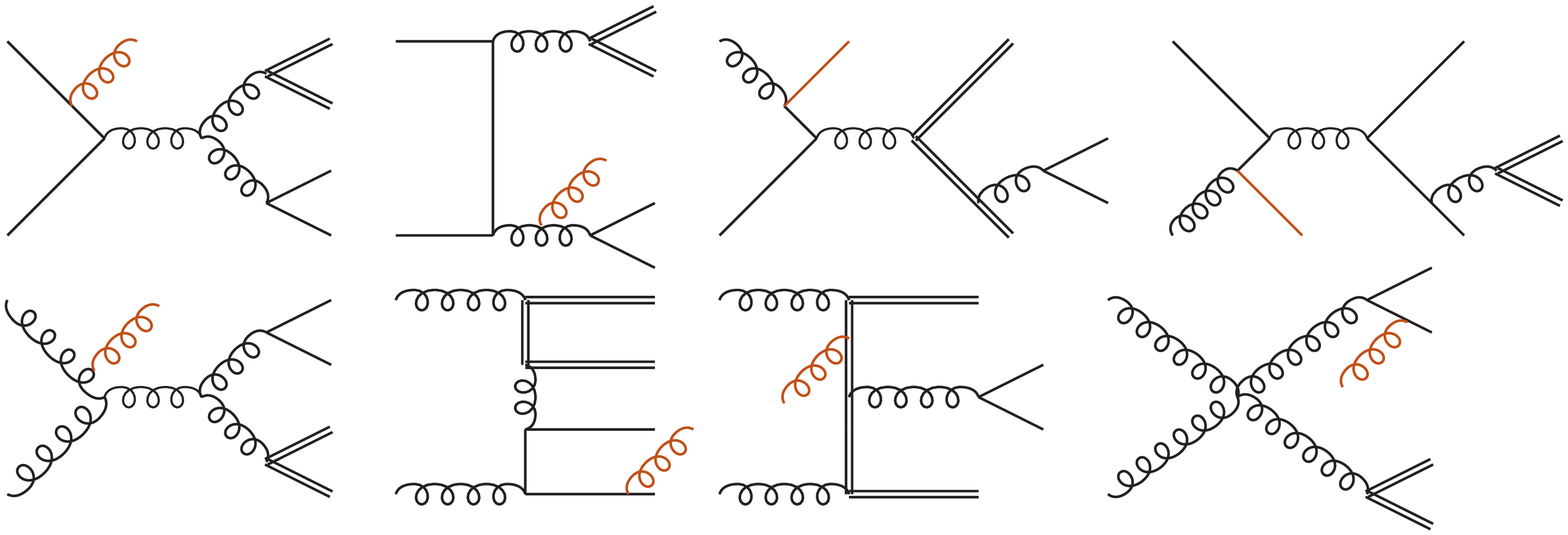}
\end{center}
\vspace{-0.2cm}
\caption{\it \label{fig:R}  A representative set of Feynman diagrams
  contributing to the real emission corrections to the $p\bar{p} \to
  t\bar{t}b\bar{b}$  process. Double lines correspond to top quarks,
  single lines to light quarks and wiggly ones to gluons.}
\end{figure}

%
At tree level, $t\bar{t}$ production in association with two b-jets
proceeds via the scattering of two gluons or two quarks.  A few
examples of LO graphs  contributing to $p\bar{p}  \to
t\bar{t}b\bar{b}$ production are shown in Figure \ref{fig:LO}.  The
virtual corrections are obtained from the interference of the sum of
all one-loop diagrams with the Born amplitude. One can classify them
into self-energy, vertex, box-type, pentagon-type and hexagon-type
corrections.  In Figure \ref{fig:V}  a few examples of  pentagon and
hexagon diagrams contributing to the virtual corrections to the
$p\bar{p} \to t\bar{t}b\bar{b}$  process are given.  And finally, the
real emission corrections to the LO process arise from tree level
amplitudes with one additional parton, an additional gluon, or a quark
anti-quark pair replacing a gluon. All possible contributions can be
divided into four subprocesses, $q\bar{q} \to t\bar{t}b\bar{b} g$, $gg
\to t\bar{t}b\bar{b}g$, $qg \to t\bar{t}b\bar{b}q$ and $gq \to
t\bar{t}b\bar{b}q$.  In Figure \ref{fig:R} a representative set of
Feynman diagrams contributing to the real emission corrections is
shown.

The calculation of NLO  corrections to $p\bar{p}\to t\bar{t}b\bar{b}$
production  proceeds along the same lines as our earlier work on  $pp
\to t\bar{t}b\bar{b}$ \cite{arXiv:0907.4723}, $pp(p\bar{p})\to
t\bar{t}jj$ \cite{arXiv:1002.4009,arXiv:1108.2851} and
$pp(p\bar{p})\to W^+W^- b̄\bar{b} \to e^+\nu_e \mu^{-} \bar{\nu}_{\mu
}b̄\bar{b}$ \cite{arXiv:1012.4230}.  The methods and internal tests
developed there have therefore been straightforwardly adapted for this
project. Let us stress here, that results for the $pp \to
t\bar{t}b\bar{b}$ and $pp(p\bar{p})\to W^+W^- b̄\bar{b}$ processes have
also been obtained by other groups
\cite{arXiv:0905.0110,arXiv:1001.4006,arXiv:1012.3975}.

To summarize briefly, off-shell methods and the OPP reduction
procedure \cite{hep-ph/0609007},  as implemented in the
\textsc{Helac-NLO} system  \cite{arXiv:1110.1499} are used in
computing the NLO QCD  corrections for the $p\bar{p}\to
t\bar{t}b\bar{b}$ processes.  The system consists of,
\textsc{CutTools}
\cite{arXiv:0711.3596,arXiv:0802.1876,arXiv:0803.3964,arXiv:0903.0356}
and \textsc{Helac-1Loop} \cite{arXiv:0903.4665}, which handle the
virtual corrections and \textsc{Helac-Dipoles} \cite{arXiv:0905.0883}
for the real emission contributions. For the phase space integration,
the \textsc{Kaleu} package \cite{arXiv:1003.4953} is used and results
are cross checked with the help of the \textsc{Phegas} phase space
generator \cite{hep-ph/0007335}.  Moreover, \textsc{OneLOop}
\cite{arXiv:1007.4716}  is employed for the evaluation of the one-loop
scalar functions. 

LO results, which are  generated with
\textsc{Helac-Dipoles} have been cross checked with
\textsc{Helac-Phegas}\footnote{Let us stress that
\textsc{Helac-Phegas} has already been extensively used and tested, see
e.g. \cite{hep-ph/0311273,arXiv:0706.2569,arXiv:0810.4861,arXiv:0912.0749,
arXiv:1106.3178}.} 
\cite{hep-ph/0002082,hep-ph/0512150,arXiv:0710.2427}.  Perfect
agreement has been found.  The numerical stability of virtual
corrections has been monitored
by checking Ward identities at every phase space point. Those events,
which violate gauge invariance ($0.13\%$ out of all $2\times 10^5$ generated
unweighted  events for which one-loop contributions are calculated) 
have  been recalculated with quadruple
precision. In addition, the cancellation of soft and 
collinear divergences after combining virtual and real corrections has
been checked numerically for a few phase space points.
%
%

\section{Phenomenological Results}

%
In the following, we present predictions for the $t\bar{t}b\bar{b} + X$
process  at the TeVatron run II with  ${\sqrt{s} = 1.96}$ TeV. We use
the Tevatron  average mass of the top quark $m_t$ = 173.3 GeV
 as measured  by the CDF and D0 experiments \cite{arXiv:1007.3178}. The
masses of all other quarks, including b quarks, are neglected.  
We have consistently employed the
MSTW2008  set of parton distribution functions (PDFs)
\cite{arXiv:0901.0002}.  In particular, we take MSTW2008LO  PDFs with
1-loop running $\alpha_s$ at LO and MSTW2008NLO PDFs with 2-loop
running $\alpha_s$ at NLO, including five active
flavors. Contributions induced  by bottom-quark densities are not taken into
account due to their negligible size. The renormalization and  
factorization scales are set to a common 
value $\mu_R = \mu_F = \mu = m_{t}$. All final-state massless partons with
pseudorapidity $|\eta| < 5$, defined as 
$\eta= -\ln \left[ \tan(\theta/2) \right]$, where $\theta$ is 
the angle between the parton momentum and the beam axis,
are  recombined into jets with a resolution parameter
$R=0.8$ via an IR-safe algorithm. We have applied three different 
jets algorithms: $k_T$ \cite{CERN-TH-6473-92,CERN-TH-6775-93,hep-ph/9305266},  
{\it anti}-$k_T$ \cite{arXiv:0802.1189} and the Cambridge/Aachen 
(C/A) algorithm \cite{hep-ph/9707323}. For our main setup, two b-jets
are required to have
\begin{equation}
p_{T}(b) > 20 ~ {\rm GeV}, ~~|y(b)| < 2.5, 
~~\Delta R_{b\bar{b}} > 0.8\,,
\end{equation}  
where $p_T(b)$, $y(b)$  are the transverse momentum and  rapidity of the b-jet,
and $\Delta R_{b\bar{b}}$ is the separation in  the plane of rapidity and
azimuthal angle between $b\bar{b}$ pairs.  Jets momenta are formed as the
four-vector sum of massless  parton momenta.  At LO, there are exactly 
two massless
final state partons, which are identified as two b-jets, provided they pass
the cuts described above. At NLO, a third parton might emerge.  It could be
recombined with another parton to give a b-jet,  or an additional jet with
unrestricted kinematics may appear.
Outgoing top and anti-top  quarks
are left on-shell, they do not undergo any cut selection.
%

\subsection{Integrated Cross Sections}

%
We start with the total cross sections. In Table  \ref{tab:tev1}
and Table \ref{tab:tev2} integrated cross sections at LO and  NLO for
$p\bar{p}\rightarrow t\bar{t}b\bar{b} ~+ X$  production  at the
TeVatron run II are presented.  In Table \ref{tab:tev2}, the scale dependence 
of the total cross section is also given. Using a  fixed scale
independent of the final state kinematics, which we set to $m_t$ and
estimate the error with the usual variation in the range between
$m_t/2$ and $2m_t$ our findings can be summarized as follows
\begin{equation}
\label{results-lo}
\sigma_{\rm LO}({\rm \textsc{TeVatron}}, ~~m_t = 173.3 ~{\rm GeV}, 
{\rm \textsc{MSTW2008lo}}~ ~) = 3.912^{~+3.496(89\%)}_{~-1.705(43\%)} 
~{\rm fb}\, , 
\end{equation}
\begin{equation}
\label{results-nlo}
\sigma_{\rm NLO}({\rm \textsc{TeVatron}}, m_t = 173.3 ~{\rm GeV}, 
{\rm \textsc{MSTW2008nlo}}) = 3.835^{~+0.992(26\%)}_{~-1.015(26\%)}  ~{\rm fb} \, ,
\end{equation}
which leaves us with an NLO ${\cal K}$-factor equal to ${\cal K} =
0.98$ and a  negative NLO QCD correction of the order of $2\%$.  This
is very different from the LHC case,  where the same cut selection
implies NLO corrections  of the order of $77\%$
\cite{arXiv:0905.0110,arXiv:0907.4723}.  There, a dynamical scale,
$\mu=m_t\sqrt{p_T(b)p_T(\bar{b})}$, and $m_{b\bar{b}} \ge 100$ GeV
had to be introduced in order to reduce the high corrections down
to $25\%-30\%$  \cite{arXiv:1001.4006}.   The dissimilarity between
the size of NLO QCD corrections   comes mostly from the difference in
the production process.  At the TeVatron, with our cut selection the
$q\bar{q}$̄ channel dominates the total LO $p\bar{p}$ cross section at
about $91\%$  followed by the gg channel with about $9\%$. In contrast,
the $gg$ channel comprises about $94\%$ of the LO $pp$ 
cross section at the LHC,  followed by the $q\bar{q}$̄ channel with about
$6\%$. 

To assess the effect of changing the jet algorithm, we compare NLO
results for different jet finders, 
the $k_T$, {\it anti}-$k_T$ and the inclusive Cambridge/Aachen
(C/A) jet  algorithms as shown in Table \ref{tab:tev1}. 
No significant change of the results due to the choice of the jet
algorithm is observed. Differences are below $1\%$.  

In Equation (\ref{results-lo}) and Equation (\ref{results-nlo}) the scale
dependence is indicated by the upper and lower values. The upper
(lower) value represents the change when the scale is shifted towards
$\mu = m_t/2 ~(\mu = 2m_t)$. Rescaling the common scale from the
default value  $m_t$ up and down by a factor $2$ changes the cross
section at LO by  $89\%$. On the other hand, the improvement  
in the scale  stability at NLO is prominent. The scale uncertainty 
is  reduced down  to $26\%$.
%
\begin{table}[h]
\begin{center}
  \begin{tabular}{c|c |c| c| c}
     & $\sigma_{\rm LO}$ [fb]       &
     $\sigma_{\rm NLO}^{\rm anti-k_T}$  [fb] &
     $\sigma_{\rm NLO}^{\rm k_T}$   [fb] &
     $\sigma_{\rm NLO}^{\rm C/A}$   [fb] \\
\hline
    $\alpha_{\rm max}=1$   & 3.912(3)   & 3.835(3)  & 3.859(3) & 3.853(3)  \\
   $\alpha_{\rm max}= 0.01 $  & 3.912(3) & 3.836(5) & 3.861(5) & 3.856(5)  \\
  \end{tabular}
\end{center}
  \caption{\it \label{tab:tev1} Integrated cross section at LO and NLO for
    $p\bar{p}\rightarrow t\bar{t}b\bar{b} ~+ X$  production  at the TeVatron
    run II.  Results for three
    different jet algorithms and two different values of $\alpha_{\rm max}$ 
    are presented. The scale choice is $\mu=m_{t}$.}
\end{table}
\begin{table}[h]
\begin{center}
  \begin{tabular}{c|c |c |c}
   & $0.5\cdot m_{t}$ & $1 \cdot m_{t}$ & $2 \cdot m_{t}$ \\
\hline
   $\sigma_{\rm LO}$ [fb]     &  7.408(5)   & 3.912(3)   &  2.207(2)     \\
   $\sigma_{\rm NLO}$ [fb]  &  4.827(8)     & 3.835(3) &   2.820(3)    \\
  \end{tabular}
\end{center}
  \caption{\it \label{tab:tev2} Scale dependence of the total cross section
    for $p\bar{p}\rightarrow t\bar{t}b\bar{b} ~+ X$  production  at the
    TeVatron run II at LO and NLO with $\mu = \xi \cdot m_{t}$.}
\end{table}
%

In addition, integrated NLO cross sections for  two values of the
unphysical cutoff parameter $\alpha_{\rm max}$, which is a 
common modification of subtraction terms in the phase space away from 
the singularity first introduced in \cite{hep-ph/9806317}, are  given 
in Table \ref{tab:tev1}. To be specific,  $\alpha_{\rm max}=1$, which
corresponds to the original  formulation of
\cite{hep-ph/9605323,hep-ph/0201036}, and $\alpha_{\rm max}=0.01$ are
considered. The independence of the final result on the value of  the
$\alpha_{\rm max}$ parameter is obtained at the per-mil level. This is
a strong consistency check of the calculation of the real emission
part. For more details on the $\alpha_{\rm max}$ implementation in the 
\textsc{Helac-Dipoles} package see e.g. \cite{arXiv:0905.0883,arXiv:0907.4723}.

Results for a slightly modified setup have also been generated,  to
determine the stability of the NLO ${\cal
  K}$-factor. On the one hand,  a higher  transverse momentum  cut on the
b-jet of $40$ GeV has been chosen.  On the other hand, a higher jet
separation cut, $\Delta R_{b\bar{b}}>1$ has been used together with a
new jet resolution parameter $R = 1$. All other parameters have been
left unchanged.  With this modified selection of cuts the integrated NLO
${\cal K}$-factor  has changed from $0.98$ to $0.88$.  
More precisely, we have obtained 
the following integrated cross section at LO and NLO 
\begin{equation}
\sigma_{\rm  LO}=0.8135(6) ~{\rm fb}\, , 
\end{equation}
\begin{equation}
\sigma_{\rm  NLO}=0.7121(10)   ~{\rm fb}\, , 
\end{equation}
which results in negative NLO QCD corrections of the order of $12\%$.
%
\begin{figure}
\begin{center}
\includegraphics[width=0.495\textwidth]{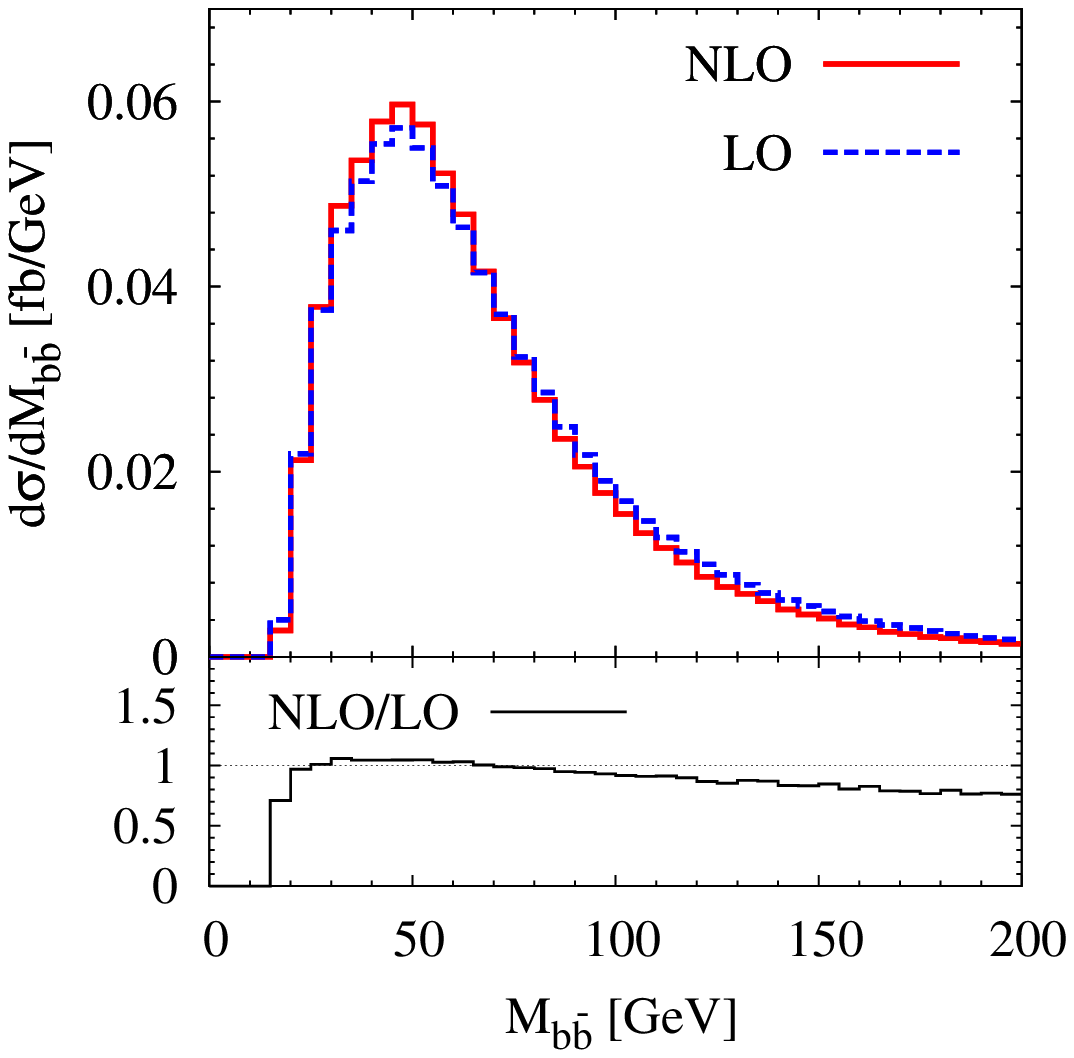}
\includegraphics[width=0.495\textwidth]{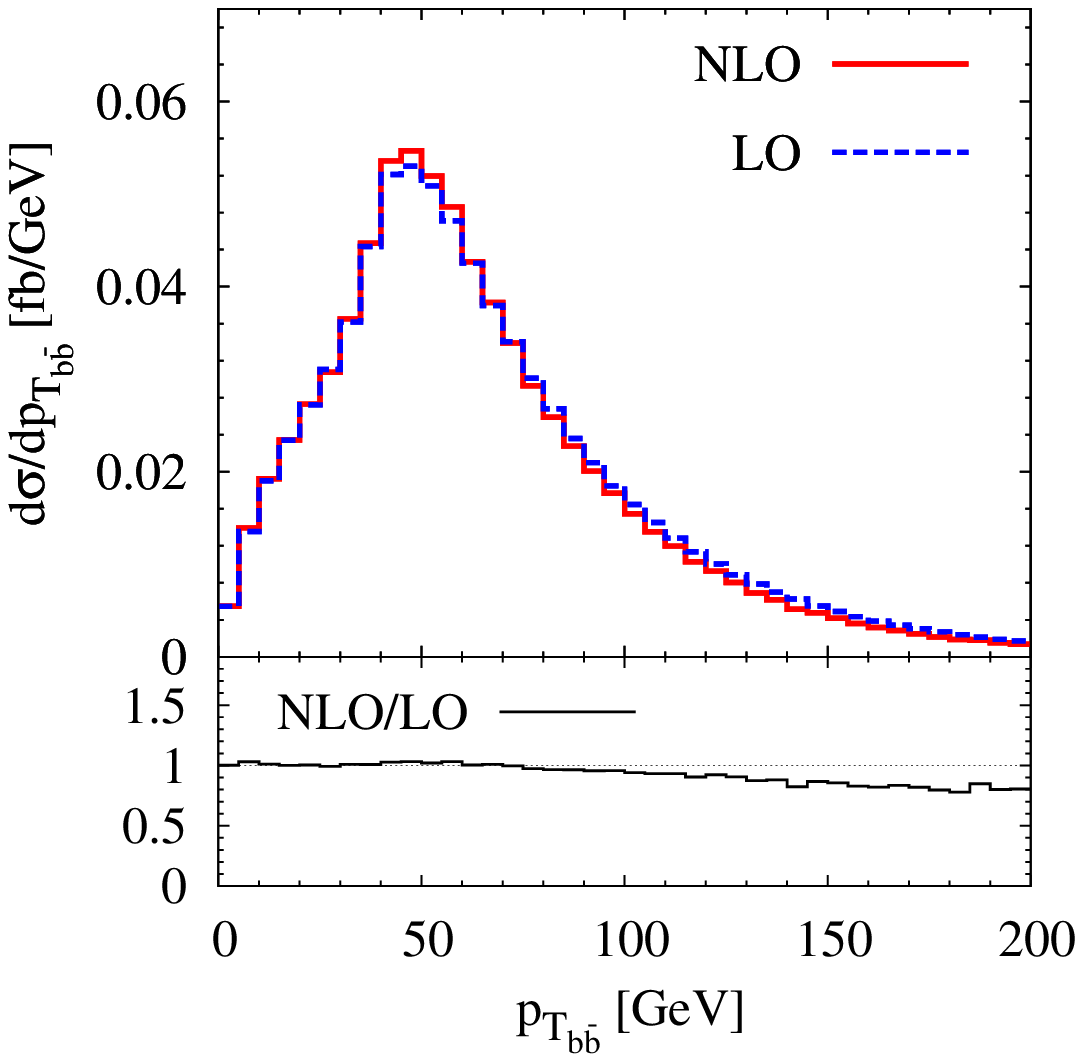}
\end{center}
\vspace{-0.2cm}
\caption{\it \label{fig:1} Distribution of the invariant mass
  $m_{b\bar{b}}$ (left panel) and  distribution in the transverse
  momentum  $p_{T_{b\bar{b}}}$  (right  panel)  of the
  bottom-anti-bottom pair for $p\bar{p}\rightarrow t\bar{t}b\bar{b} +
  X$ at the TeVatron run II at LO (blue dashed line) and NLO (red
  solid line). The lower panels display the differential ${\cal K}$
  factor.}
\end{figure}
\begin{figure}
\begin{center}
\includegraphics[width=0.495\textwidth]{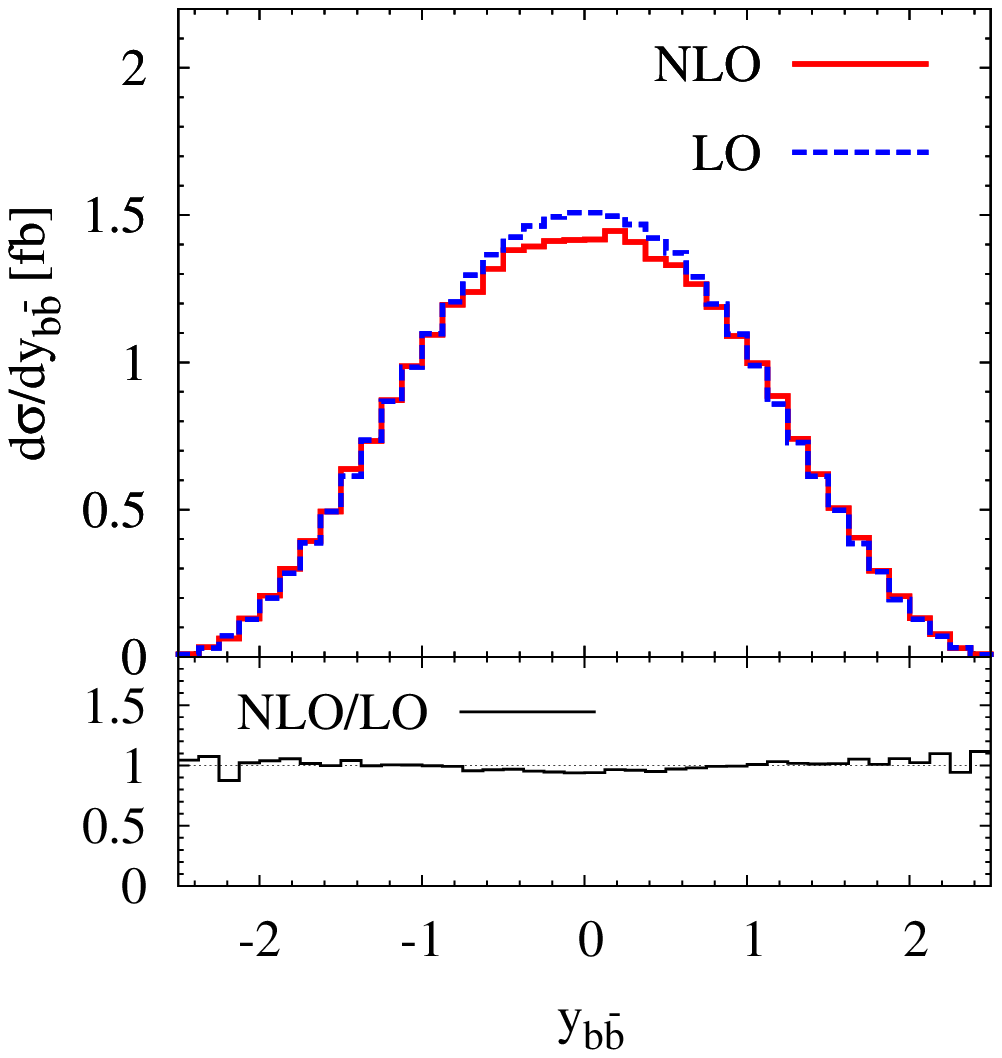}
\includegraphics[width=0.495\textwidth]{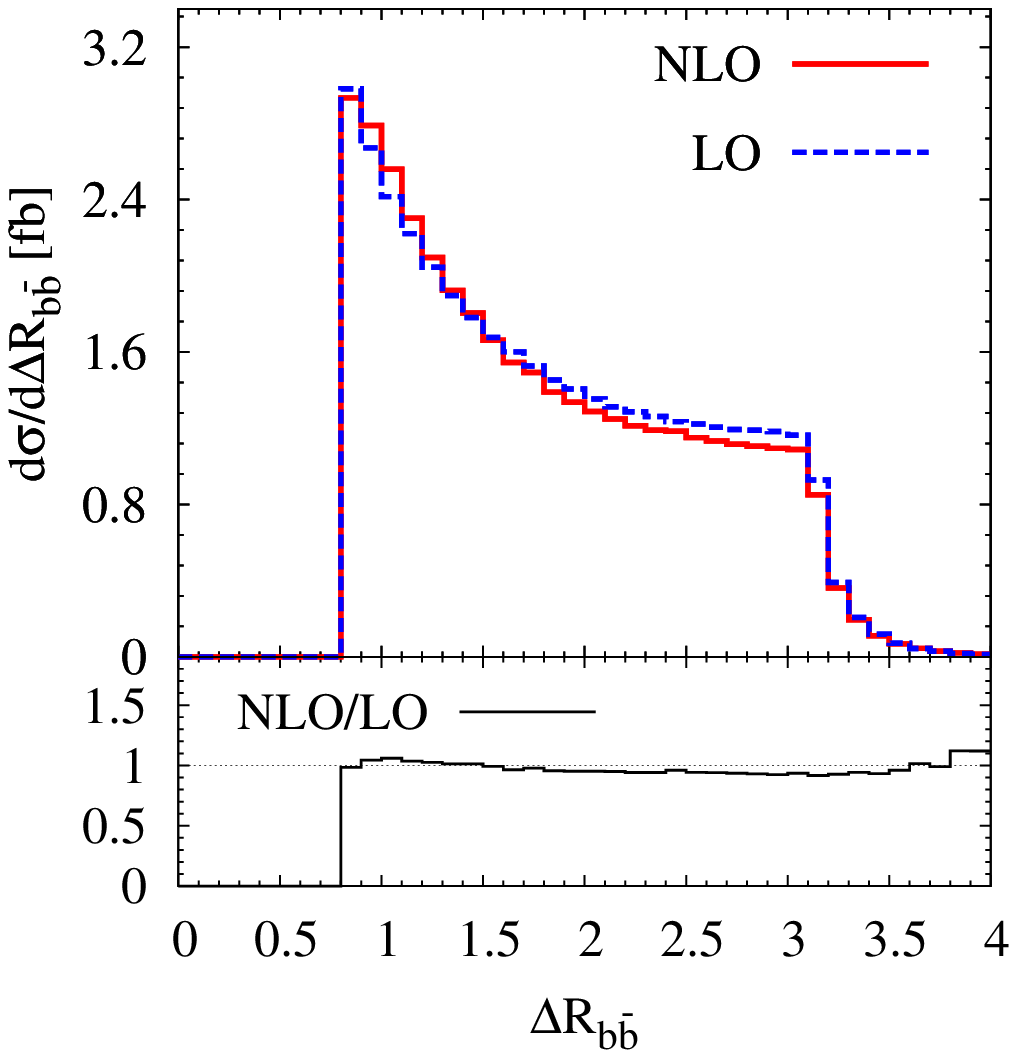}
\end{center}
\vspace{-0.2cm}
\caption{\it \label{fig:2} Distribution in the rapidity $y_{b\bar{b}}$
  (left panel) of the bottom-anti-bottom pair and  distribution of the
  $\Delta R_{b\bar{b}}$  separation (right panel) for
  $p\bar{p}\rightarrow t\bar{t}b\bar{b} + X$ at the TeVatron run II at
  LO (blue dashed line) and NLO (red solid line). The lower panels
  display the differential ${\cal K}$ factor.}
\end{figure}
\begin{figure}
\begin{center}
\includegraphics[width=0.495\textwidth]{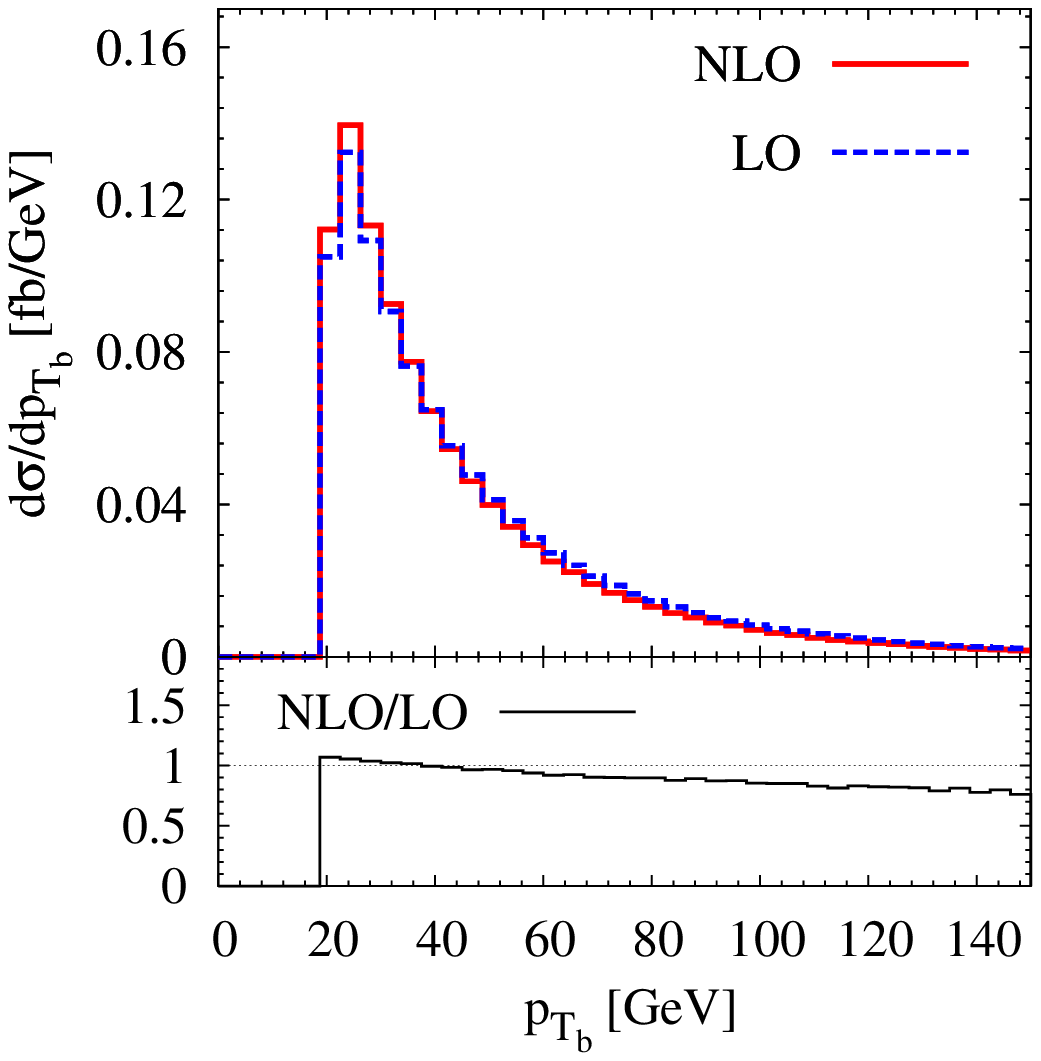}
\includegraphics[width=0.495\textwidth]{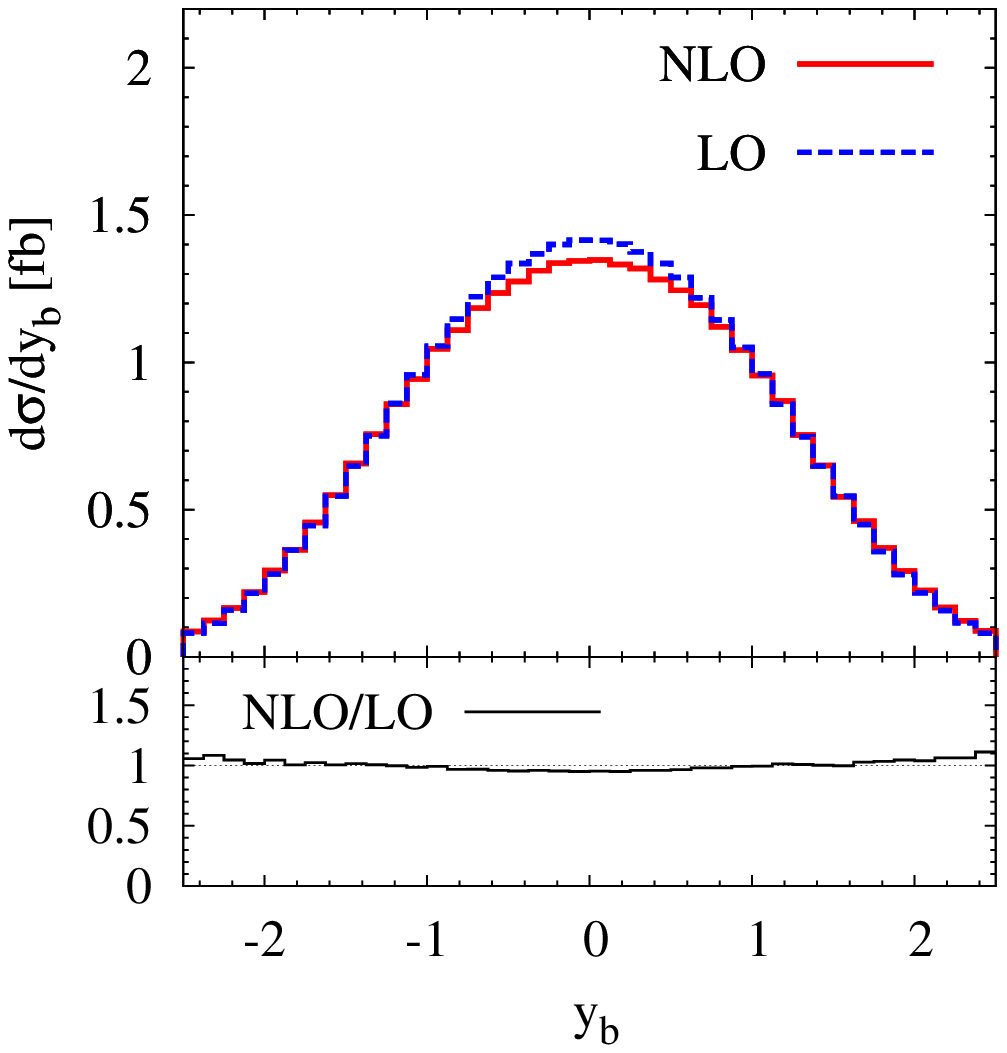}
\end{center}
\vspace{-0.2cm}
\caption{\it \label{fig:3} Distribution in the
  transverse momentum $p_{T_{b}}$  (left panel) and distribution
  in the  rapidity $y_{b}$ (right panel) of the  bottom quark
  for $p\bar{p}\rightarrow t\bar{t}b\bar{b} + X$ at the TeVatron run
  II at LO (blue dashed line) and NLO (red solid line). The lower
  panels display the differential ${\cal K}$ factor.}
\end{figure}
%

\subsection{Differential Cross Sections}

%
In the following, the impact of QCD corrections  on   differential
cross sections is analyzed. The differential  distributions
relevant for  Higgs boson searches in the  $t\bar{t}H\to
t\bar{t}b\bar{b}$ channel are plotted first.  In Figure \ref{fig:1}, the
distribution of the invariant mass $m_{b\bar{b}}$ and the distribution in
the transverse momentum  $p_{T_{b\bar{b}}}$  of the bottom-anti-bottom
pair  for $p\bar{p}\rightarrow t\bar{t}b\bar{b} + X$ at the TeVatron
run II is plotted.  The dashed curve corresponds to the LO,
whereas the solid one to the NLO  result. The upper panels show the
distributions themselves while the lower  panels display the ratio of
the NLO value to the LO result, calculated according to
\begin{equation}
{\cal K}({\cal O}) = 
\frac{d\sigma_{\rm NLO}/d{\cal O}}{ d\sigma_{\rm LO}/d{\cal O}} \, ,
\end{equation}
for an observable ${\cal O}$ under investigation,   called the differential or
dynamical ${\cal K}$ factor. In Figure \ref{fig:2}, the distribution in the
rapidity $y_{b\bar{b}}$ of the bottom-anti-bottom pair and the distribution of
the $\Delta R_{b\bar{b}}$  separation between two b-jets are shown.  In all
cases, the small NLO corrections of the integrated cross section are 
also visible
at the differential level. In case of the most important observable, the
$m_{b\bar{b}}$ differential  distribution, in the phenomenologically relevant
region, i.e. below $140$ GeV, corrections of the order of $\sim 10\%$ are
reached.  Besides, the separation between the b-jets of $0.8$ together with
the transverse momentum cut on b-jets of $20$ GeV sets an effective lower
bound on the invariant mass of two b-jets of the order of $m_{b\bar{b}} \gtrsim
15.6$ GeV \cite{arXiv:0907.4723}. If we had chosen $\Delta R_{b\bar{b}} \ge
0.5$ the minimum  $m_{b\bar{b}}$ would have been around $9.9$ GeV instead. 

In view of the small NLO corrections to the total cross 
sections as well as to the differential distributions presented here, 
we conclude that a meaningful analysis at the TeVatron can be
performed with
the present setup, i.e. with the renormalization and factorization
scales fixed to a common value (the mass of top-quark).

Finally, for completeness   the distributions
for b-jet kinematics are given in Figure \ref{fig:3}.  
Namely, the distribution in the
transverse momentum $p_{T_{b}}$ and the distribution in the  rapidity
$y_{b}$  of the  bottom quark are shown. Similarly to the
bottom-anti-bottom pair kinematics, also here,  small NLO corrections
are visible. The shape of kinematic distributions can be distorted by
$20\%$ at most in some regions  of phase space.  This is once again 
contrary to the
LHC case, where  large and relatively constant NLO corrections have
been obtained for $\mu=m_t$, whereas  a dynamical scale and
additional cuts allowed to reduce them down to $20\%-40\%$.
%

\subsection{Forward-Backward Asymmetry}

%
\begin{figure}
\includegraphics[width=0.495\textwidth]{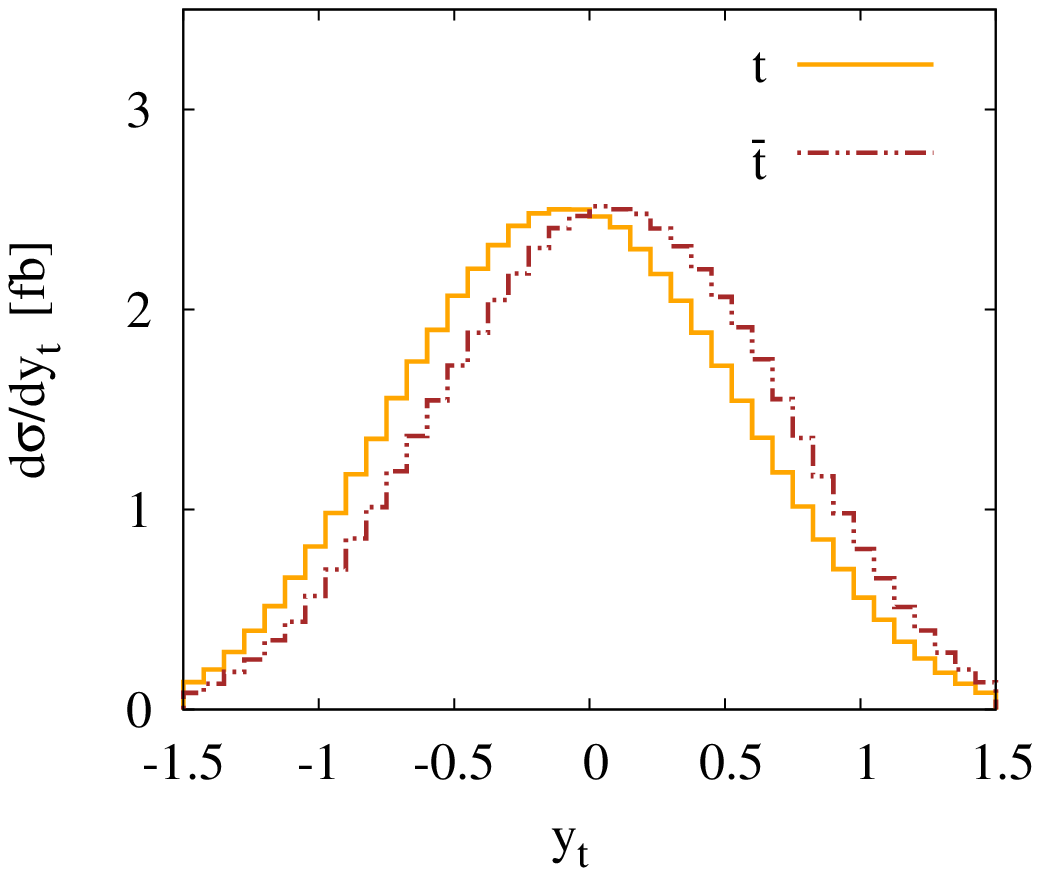}
\includegraphics[width=0.495\textwidth]{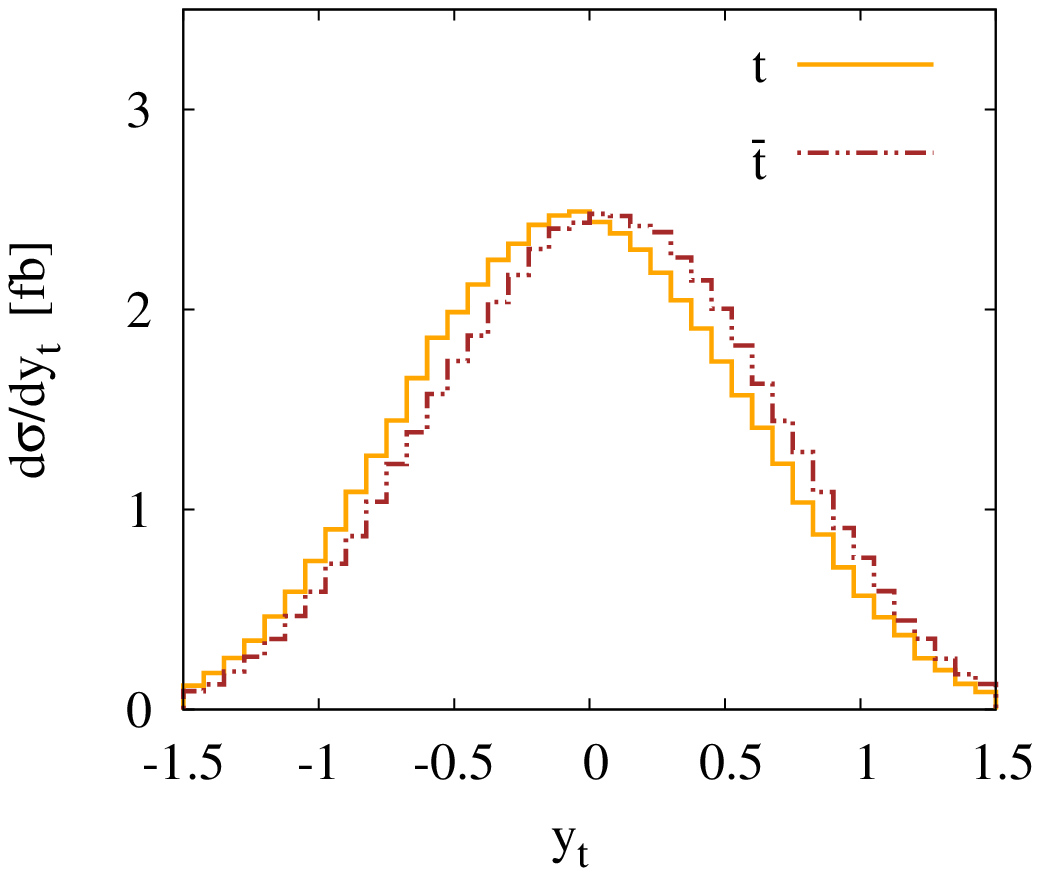}
\caption{\it \label{tev:asymmetry1} 
  Differential cross section distributions as a function of
  rapidity, $y_t$, of  the top and anti-top quark at LO (left
  panel) and NLO (right panel)  for  $ p\bar{p} \to t
  \bar{t} b\bar{b} + X$ production  at the TeVatron run II.  
  The (orange) solid curve corresponds to the top
  quark, whereas the (brown) dash-dotted  one to the anti-top quark.}
\end{figure}
\begin{figure}
\includegraphics[width=0.495\textwidth]{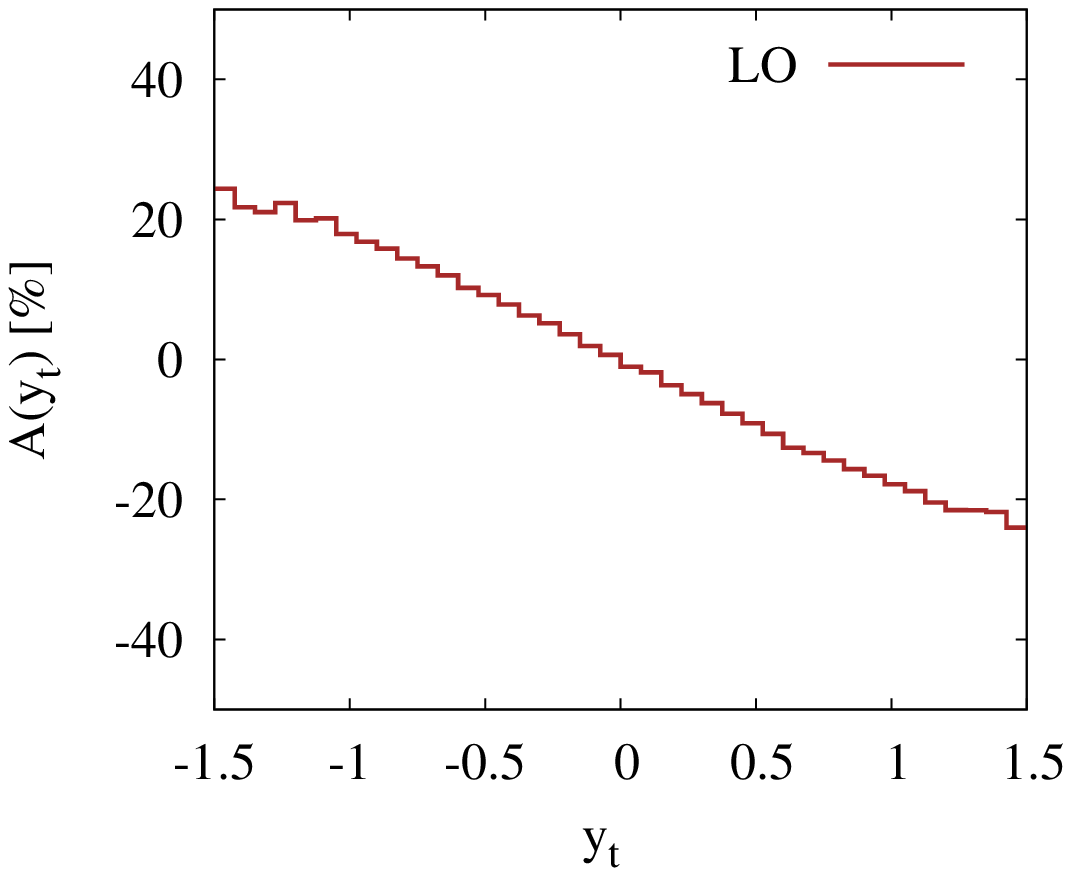}
\includegraphics[width=0.495\textwidth]{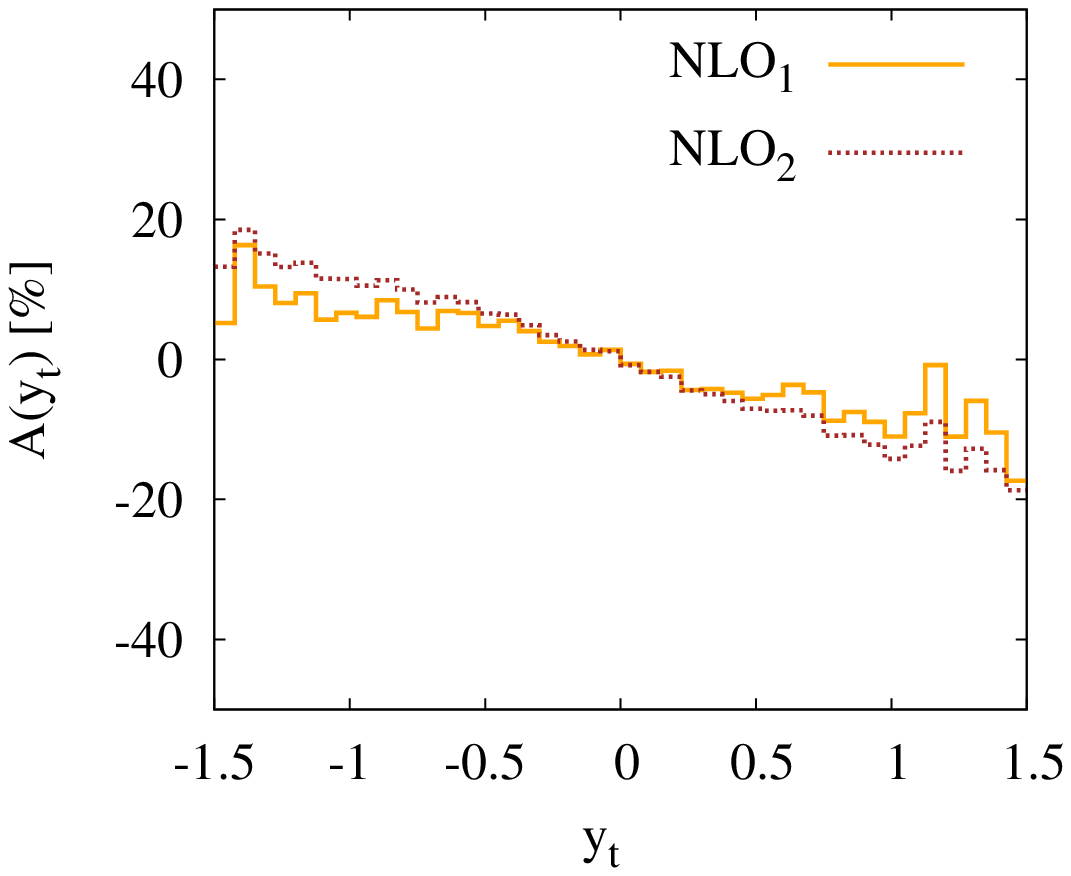}
\caption{\it \label{tev:asymmetry2}  Differential  asymmetry
  ${\cal A}(t)$, as a function of the top quark rapidity at LO (left
  panel) and NLO (right panel) for  $ p\bar{p} \to t \bar{t} b\bar{b}
  + X$ production  at the TeVatron run II. $NLO_1$ refers to a result
  with a  consistent expansion in $\alpha_s$, while $NLO_2$ to the
  unexpanded one.}
\end{figure}
%
As a bonus of our study, we calculated the integrated top quark 
forward-backward asymmetry
for the $t\bar{t}b\bar{b}$ production process at the TeVatron. 
At LO the  asymmetry is defined as 
\begin{equation}
{\cal A}_{\rm FB, LO}^{\rm t} =\frac{\sigma_{\rm LO}(y_t > 0)  - \sigma_{\rm
  LO}(y_t < 0)}{\sigma_{\rm LO}(y_t > 0)  + \sigma_{\rm
  LO}(y_t < 0)} \, ,
\end{equation}
where $y_t$ is the rapidity of the top quark and 
$\sigma_{\rm LO}^{\pm} = \sigma_{\rm LO}(y_t > 0)  \pm \sigma_{\rm
  LO}(y_t < 0)$ is evaluated with LO PDFs and LO $\alpha_s$. 
On the other hand, the asymmetry at NLO is expressed through 
\begin{equation}
\label{eq:nlo}
{\cal A}_{\rm FB, NLO}^{\rm t}  = \frac{\sigma_{\rm LO}^{-}+\delta \sigma_{\rm
  NLO}^{-}}{\sigma_{\rm LO}^{+}+\delta \sigma_{\rm  NLO}^{+}} \, ,
\end{equation}
where $\delta \sigma_{\rm NLO}^{\pm}$ are the NLO contributions to the
cross sections and $\sigma_{\rm LO}^{\pm}$ are evaluated  this time
with NLO PDFs and NLO $\alpha_s$.  The ratio  generates contributions
of ${\cal O}(\alpha_s^2)$ and higher, which are affected by
unknown next-to-next-to-leading order contributions. Therefore,    it
is necessary to expand Equation (\ref{eq:nlo}) to first order in
$\alpha_s$.  The following definition is then obtained
\cite{hep-ph/0703120,arXiv:0810.0452}
\begin{equation}
{\cal A}_{\rm FB, NLO}^{\rm t}  = \frac{\sigma_{\rm LO}^{-}}{\sigma_{\rm LO}^{+}}
\left(  1 + \frac{\delta\sigma_{\rm NLO}^{-}}{\sigma_{\rm LO}^{-}} -
\frac{\delta \sigma_{\rm NLO}^{+}}{\sigma_{\rm LO}^{+}} \right) \, .
\end{equation}

The integrated forward-backward asymmetry of the top quark at LO  for 
$t\bar{t}b\bar{b}$ production amounts to (the error in parentheses
corresponds to scale variation)
\begin{equation}
\label{eq:lo}
{\cal A}_{\rm FB, LO}^{\rm t}  = -0.088(2)   \, .
\end{equation}
With NLO corrections, the asymmetry is reduced down to  
\begin{equation}
\label{eq:nloexp}
{\cal A}_{\rm FB, NLO}^{\rm t}  = -0.044(6)\, ,
\end{equation}
when a definition with a consistent expansion  in $\alpha_s$ is used.
For an unexpanded ratio of the NLO cross sections, the result is
\begin{equation}
{\cal A}_{\rm FB, NLO}^{\rm t}  = -0.061(14)\, .
\end{equation}
The two NLO definitions give results which differ by about $40\%$ for
the central scale.   Also the theoretical error as calculated
from the scale dependence is more than a factor $2$ higher in the
latter case.

In Figure \ref{tev:asymmetry1}, the rapidity distributions for the top and
anti-top quarks are presented  at LO and NLO. Results are not
symmetric around $y_t=0$ and are shifted  to  a forward direction  for
the anti-top quarks and  a backward direction for the top quarks.
This shows that anti-top quarks are preferentially emitted in the
direction  of the incoming protons.

In Figure \ref{tev:asymmetry2}, we have also plotted the  differential
asymmetry, ${\cal A}(y_t)$. It rises up to $\pm 20\%$  at LO and well
above  $\pm 10\%$ at NLO in suitably chosen kinematical regions. After
including the   NLO corrections the forward-backward asymmetry of top
quarks is reduced by a factor 2 as can be seen both from 
Equation (\ref{eq:lo}) and  Equation (\ref{eq:nloexp})  
as well as  in Figure \ref{tev:asymmetry2}.
%

\section{Summary and Conclusions}

%
In this paper, a computation of the NLO QCD corrections  to the top
quark pair production in association with two hard b-jets at  the
TeVatron run II has been presented. The total cross section and its
scale dependence,  together with a few differential distributions have
been given. The impact  of the NLO QCD corrections on the integrated cross
sections  as calculated for the fixed scale $\mu_F=\mu_R=\mu=m_t$,  is
small, of the  order of $2\%$. Moreover, the NLO QCD corrections
reduce the scale uncertainty  of the total cross sections as compared
to LO calculations. As a further matter,  the NLO corrections to the
differential distributions are below $10\%$ in the phenomenologically
significant regions, reaching $20\%$ at some corners of the phase
space.  

Since at the TeVatron, the corrections are small,  the integrated and
differential ${\cal K}$-factors do not necessarily need to be  applied
in the background estimation  for the $t\bar{t}H\to t\bar{t}b\bar{b}$
signal process at the TeVatron.  At least not for the observables
which have been scrutinized here. Definitely, an
application of the large LHC ${\cal K}$-factor to TeVatron
phenomenological analyzes dramatically overestimates  the size of 
the irreducible $t\bar{t}b\bar{b}$ background. 

Let us conclude by noting that the predicted rates are very
small. Combined with the luminosity of the TeVatron they amount to
just a few events. The relevance of the present study lies, however,
in the fact that Higgs exclusion bounds are obtained using a
combination of data from different production channels with the $t\bar
tH$ process, and thus its irreducible  background, also taken into
account by the {\textsc CDF} and {\textsc D0} collaborations.
%

\acknowledgments 

%
The calculations have been performed on the Grid Cluster of the Bergische 
Universit\"at Wuppertal, financed by the Helmholtz - Alliance 
“Physics at the Terascale” and the BMBF. The
author was supported  by the Initiative and Networking Fund of the 
Helmholtz Association, contract HA-101 (“Physics at the Terascale”).

The author would like to thank Christian Schwanenberger for motivating her 
to perform this study.

\providecommand{\href}[2]{#2}\begingroup\raggedright\endgroup


\begin{thebibliography}{10}

\bibitem{hep-ph/0107081}
W.~Beenakker, S.~Dittmaier, M.~Kramer, B.~Plumper, M.~Spira and P.~M.~Zerwas, 
{\it Higgs radiation off top quarks at the Tevatron and the LHC}, 
Phys.\ Rev.\ Lett.\ \ {\bf 87} (2001) 201805, 
{\tt [hep-ph/0107081]}.

\bibitem{hep-ph/0107101}
L.~Reina and S.~Dawson, 
{\it Next-to-leading order results for t anti-t h production at the Tevatron}, 
Phys.\ Rev.\ Lett.\ \ {\bf 87} (2001) 201804, 
{\tt [hep-ph/0107101]}.

\bibitem{hep-ph/0109066}
L.~Reina, S.~Dawson and D.~Wackeroth, 
{\it QCD corrections to associated t anti-t h production at the Tevatron},
Phys.\ Rev.\ D\ {\bf 65} (2002) 053017, {\tt [hep-ph/0109066]}.

\bibitem{hep-ph/0211352}
W.~Beenakker, S.~Dittmaier, M.~Kramer, B.~Plumper, M.~Spira and P.~M.~Zerwas, 
{\it NLO QCD corrections to t anti-t H production in hadron collisions}, 
Nucl.\ Phys.\ B\ {\bf 653} (2003) 151, {\tt [hep-ph/0211352]}.

\bibitem{arXiv:1107.5518}
CDF and D0 Collaborations, 
{\it Combined CDF and D0 Upper Limits on Standard Model 
Higgs Boson Production with up to 8.6 $fb^{-1}$ of Data},
{\tt arXiv:1107.5518 [hep-ex]}.

\bibitem{D0}
D0 Collaboration, {\it Search for the Standard Model Higgs boson in the 
$t\bar{t}H \to t\bar{t}b\bar{b}$ channel}, {\tt D0 note 5739-CONF}, (2009).

\bibitem{CDF}
CDF Collaboration, {\it Search for SM Higgs boson production in association 
with $t\bar{t}$ using no lepton final state}, {\tt CDF note 10582}, (2011).

\bibitem{arXiv:0907.4723}
G. Bevilacqua, M. Czakon, C. G. Papadopoulos, R. Pittau and M. Worek,
{\it Assault on the NLO Wishlist: $pp \to t\bar{t}b\bar{b}$},  
JHEP\ {\bf 0909} (2009) 109, {\tt [arXiv:0907.4723 [hep-ph]]}.

\bibitem{arXiv:1002.4009}
G. Bevilacqua, M. Czakon, C. G. Papadopoulos and M. Worek, 
{\it Dominant QCD Backgrounds in Higgs Boson Analyses at the LHC: A Study of 
pp $\to$ t anti-t + 2 jets at Next-To-Leading Order}, 
Phys.\ Rev.\ Lett.\ \ {\bf 104} (2010) 162002, {\tt [arXiv:1002.4009 [hep-ph]]}.

\bibitem{arXiv:1108.2851}
G. Bevilacqua, M. Czakon, C. G. Papadopoulos and M. Worek, 
{\it Hadronic top-quark pair production in association with two jets at 
Next-to-Leading Order QCD}, 
{\tt arXiv:1108.2851 [hep-ph]}.

\bibitem{arXiv:1012.4230}
G. Bevilacqua, M. Czakon, A. van Hameren, C. G. Papadopoulos and M. Worek, 
{\it Complete off-shell effects in top quark pair hadroproduction with leptonic 
decay at next-to-leading 
order},  JHEP\ {\bf 1102} (2011) 083, {\tt [arXiv:1012.4230 [hep-ph]]}.

\bibitem{arXiv:0905.0110}
A. Bredenstein, A. Denner, S. Dittmaier and S. Pozzorini, 
{\it NLO QCD corrections to pp $\to$ t anti-t b anti-b + X at the LHC}, 
Phys.\ Rev.\ Lett.\ \ {\bf 103} (2009) 012002, {\tt [arXiv:0905.0110 [hep-ph]]}.

\bibitem{arXiv:1001.4006}
A. Bredenstein, A. Denner, S. Dittmaier and S. Pozzorini, 
{\it NLO QCD Corrections to Top Anti-Top Bottom Anti-Bottom Production at the 
LHC: 2. full hadronic results},
JHEP\ {\bf 1003} (2010) 021, {\tt [arXiv:1001.4006 [hep-ph]]}.

\bibitem{arXiv:1012.3975}
A. Denner, S. Dittmaier, S. Kallweit and S. Pozzorini, 
{\it NLO QCD corrections to WWbb 
production at hadron colliders}, Phys. Rev. Lett. {\bf 106} (2011) 052001, 
{\tt [arXiv:1012.3975 [hep-ph]]}.

\bibitem{hep-ph/0609007}
G.~Ossola, C.~G.~Papadopoulos and R.~Pittau, {\it Reducing full one-loop 
amplitudes to scalar integrals at the integrand level},  
Nucl.\ Phys.\ B\ {\bf 763} (2007) 147,  {\tt [hep-ph/0609007]}.

\bibitem{arXiv:1110.1499}
G.~Bevilacqua, M.~Czakon, M.~V.~Garzelli, A.~van Hameren, A.~Kardos, 
C.~G.~Papadopoulos, R.~Pittau and M.~Worek, {\it Helac-nlo}, 
{\tt arXiv:1110.1499 [hep-ph]}.

\bibitem{arXiv:0711.3596}
G.~Ossola, C.~G.~Papadopoulos and R.~Pittau, 
{\it CutTools: A Program implementing the OPP reduction method to compute 
one-loop amplitudes}, JHEP\ {\bf 0803} (2008) 042, 
{\tt [arXiv:0711.3596 [hep-ph]]}.

\bibitem{arXiv:0802.1876}
G.~Ossola, C.~G.~Papadopoulos and R.~Pittau, {\it On the Rational Terms of 
the one-loop amplitudes},  JHEP\ {\bf 0805} (2008) 004, 
{\tt [arXiv:0802.1876 [hep-ph]]}.

\bibitem{arXiv:0803.3964}
P.~Mastrolia, G.~Ossola, C.~G.~Papadopoulos and R.~Pittau, 
{\it Optimizing the Reduction of One-Loop Amplitudes}, 
JHEP\ {\bf 0806} (2008) 030, {\tt [arXiv:0803.3964 [hep-ph]]}.

\bibitem{arXiv:0903.0356}
P.~Draggiotis, M.~V.~Garzelli, C.~G.~Papadopoulos and R.~Pittau, 
{\it Feynman Rules for the Rational Part of the QCD 1-loop amplitudes}, 
JHEP\ {\bf 0904} (2009) 072, {\tt [arXiv:0903.0356 [hep-ph]]}.

\bibitem{arXiv:0903.4665}
A.~van Hameren, C.~G.~Papadopoulos and R.~Pittau, 
{\it Automated one-loop calculations: A Proof of concept}, 
JHEP\ {\bf 0909} (2009) 106, {\tt [arXiv:0903.4665 [hep-ph]]}.

\bibitem{arXiv:0905.0883}
M.~Czakon, C.~G.~Papadopoulos and M.~Worek, {\it Polarizing the Dipoles}, 
JHEP\ {\bf 0908} (2009) 085, {\tt [arXiv:0905.0883 [hep-ph]]}.

\bibitem{arXiv:1003.4953}
A.~van Hameren, 
{\it Kaleu: a general-purpose parton-level phase space generator}, 
{\tt  arXiv:1003.4953 [hep-ph]}.

\bibitem{hep-ph/0007335}
C.~G.~Papadopoulos, 
{\it PHEGAS: A Phase space generator for automatic cross-section computation}, 
Comput.\ Phys.\ Commun.\ \ {\bf 137} (2001) 247, {\tt [hep-ph/0007335]}.

\bibitem{arXiv:1007.4716}
A.~van Hameren, 
{\it OneLOop: For the evaluation of one-loop scalar functions},
Comput.\ Phys.\ Commun.\ \ {\bf 182} (2011) 2427, 
{\tt [arXiv:1007.4716 [hep-ph]]}.

\bibitem{hep-ph/0002082}
A.~Kanaki and C.~G.~Papadopoulos, 
{\it HELAC: A Package to compute electroweak helicity amplitudes}, 
Comput.\ Phys.\ Commun.\ \ {\bf 132} (2000) 306, {\tt [hep-ph/0002082]}.

\bibitem{hep-ph/0512150}
C.~G.~Papadopoulos and M.~Worek, 
{\it Multi-parton cross sections at hadron colliders}, 
Eur.\ Phys.\ J.\ C\ {\bf 50} (2007) 843, {\tt [hep-ph/0512150]}.

\bibitem{arXiv:0710.2427}
A.~Cafarella, C.~G.~Papadopoulos and M.~Worek, 
{\it Helac-Phegas: A Generator for all parton level processes}, 
Comput.\ Phys.\ Commun.\ \ {\bf 180} (2009) 1941, 
{\tt [arXiv:0710.2427 [hep-ph]]}.

\bibitem{hep-ph/0311273}
T.~Gleisberg, F.~Krauss, C.~G.~Papadopoulos, A.~Schaelicke and S.~Schumann,
{\it Cross-sections for multiparticle final states at a linear collider},
Eur.\ Phys.\ J.\ C {\bf 34} (2004) 173, {\tt [hep-ph/0311273]}.

\bibitem{arXiv:0706.2569}
J.~Alwall, S.~Hoche, F.~Krauss, N.~Lavesson, L.~Lonnblad, F.~Maltoni,
M.~L.~Mangano and M.~Moretti {\it et al.}, 
{\it Comparative study of various algorithms for the merging of parton showers 
and matrix elements in hadronic collisions}, Eur.\ Phys.\ J.\ C {\bf 53}
(2008) 473, {\tt [arXiv:0706.2569 [hep-ph]]}.

\bibitem{arXiv:0810.4861}
C.~Englert, B.~Jager, M.~Worek and D.~Zeppenfeld, 
{\it Observing Strongly Interacting Vector Boson Systems at the CERN Large
  Hadron Collider},  Phys.\ Rev.\ D {\bf 80} (2009) 035027,  
{\tt [arXiv:0810.4861 [hep-ph]]}.

\bibitem{arXiv:0912.0749}
S.~Actis {\it et al.}  [Working Group on Radiative Corrections and Monte Carlo 
Generators for Low Energies Collaboration], {\it Quest for precision in
hadronic crosssections at low energy: Monte Carlo tools vs. experimental data},
Eur.\ Phys.\ J.\ C {\bf 66} (2010) 585, {\tt [arXiv:0912.0749 [hep-ph]]}.

\bibitem{arXiv:1106.3178}
C.~Carloni Calame, H.~Czyz, J.~Gluza, M.~Gunia, G.~Montagna, O.~Nicrosini,
F.~Piccinini and T.~Riemann {\it et al.}, {\it NNLO leptonic and hadronic
corrections to Bhabha scattering and luminosity monitoring at meson factories},
JHEP {\bf 1107} (2011) 126, {\tt [arXiv:1106.3178 [hep-ph]]}.

\bibitem{arXiv:1007.3178}
CDF and D0 Collaborations, 
{\it Combination of CDF and D0 Results on the Mass of the Top Quark Using Up to 
5.6 $fb^{-1}$ of Data},  {\tt arXiv:1007.3178 [hep-ex]}.

\bibitem{arXiv:0901.0002}
A.~D.~Martin, W.~J.~Stirling, R.~S.~Thorne and G.~Watt, 
{\it Parton distributions for the LHC},  
Eur.\ Phys.\ J.\ C\ {\bf 63} (2009) 189, {\tt [arXiv:0901.0002 [hep-ph]]}.

\bibitem{CERN-TH-6473-92}
S.~Catani, Y.~L.~Dokshitzer and B.~R.~Webber, 
{\it The $K_{\perp}$ 
perpendicular clustering algorithm for jets in deep inelastic 
scattering and hadron collisions}, Phys.\ Lett.\ B\ {\bf 285} (1992) 291.

\bibitem{CERN-TH-6775-93}
S.~Catani, Y.~L.~Dokshitzer, M.~H.~Seymour and B.~R.~Webber, 
{\it Longitudinally invariant $K_t$ clustering algorithms for hadron hadron 
collisions}, Nucl.\ Phys.\ B\ {\bf 406} (1993) 187.

\bibitem{hep-ph/9305266}
S.~D.~Ellis and D.~E.~Soper, 
{\it Successive combination jet algorithm for hadron collisions}, 
Phys.\ Rev.\ D\ {\bf 48} (1993) 3160 {\tt [hep-ph/9305266]}.

\bibitem{arXiv:0802.1189}
M.~Cacciari, G.~P.~Salam and G.~Soyez, 
{\it The Anti-k(t) jet clustering algorithm}, 
JHEP\ {\bf 0804} (2008) 063, {\tt [arXiv:0802.1189 [hep-ph]]}.

\bibitem{hep-ph/9707323}
Y.~L.~Dokshitzer, G.~D.~Leder, S.~Moretti and B.~R.~Webber, 
{\it Better jet clustering algorithms}, JHEP\ {\bf 9708} (1997) 001, 
{\tt [hep-ph/9707323]}.

\bibitem{hep-ph/9806317}
Z.~Nagy and Z.~Trocsanyi, {\it Next-to-leading order calculation of four 
jet observables in electron positron annihilation}, 
Phys.\ Rev.\ D\ {\bf 59} (1999) 014020, 
[Erratum-ibid.\ D\ {\bf 62} (2000) 099902], {\tt [hep-ph/9806317]}.

\bibitem{hep-ph/9605323}
S.~Catani and M.~H.~Seymour, 
{\it A General algorithm for calculating jet cross-sections in NLO QCD},
Nucl.\ Phys.\ B\ {\bf 485} (1997) 291, [Erratum-ibid. B {\bf 510} (1998) 503], 
{\tt [hep-ph/9605323]}.

\bibitem{hep-ph/0201036}
S.~Catani, S.~Dittmaier, M.~H.~Seymour and Z.~Trocsanyi, 
{\it The Dipole formalism for next-to-leading order QCD calculations with
 massive partons},  Nucl.\ Phys.\ B\ {\bf 627} (2002) 189, 
{\tt [hep-ph/0201036]}.

\bibitem{hep-ph/0703120}
S.~Dittmaier, P.~Uwer and S.~Weinzierl, 
{\it NLO QCD corrections to t anti-t + jet production at hadron colliders}, 
Phys.\ Rev.\ Lett.\  {\bf 98} (2007) 262002, {\tt [hep-ph/0703120 [hep-ph]]}.

\bibitem{arXiv:0810.0452}
S.~Dittmaier, P.~Uwer and S.~Weinzierl, 
{\it Hadronic top-quark pair production in association with a hard jet at 
next-to-leading order QCD: Phenomenological studies for the Tevatron and the 
LHC},  Eur.\ Phys.\ J.\ C {\bf 59} (2009) 625, {\tt [arXiv:0810.0452 [hep-ph]]}.


\end{thebibliography}
\end{document}